# Intra- and Inter-Fraction Relative Range Verification in Heavy-Ion Therapy Using Filtered Interaction Vertex Imaging

Devin Hymers[1], Eva Kasanda[1], Vinzenz Bildstein[1], Joelle Easter[1], Andrea Richard[2,3], Artemis Spyrou[2], Cornelia Höhr[4], Dennis Mücher[1,4]

[1]Department of Physics, University of Guelph, Guelph, ON, Canada
[2]National Superconducting Cyclotron Laboratory, Michigan State University, East Lansing, MI, USA
[3]Lawrence Livermore National Laboratory, Livermore, CA, USA
[4]TRIUMF, Vancouver, BC, Canada

**Abstract**

Heavy-ion therapy, particularly using scanned (active) beam delivery, provides a precise and highly conformal dose distribution, with maximum dose deposition for each pencil beam at its endpoint (Bragg peak), and low entrance and exit dose. To take full advantage of this precision, robust range verification methods are required; these methods ensure that the Bragg peak is positioned correctly in the patient and the dose is delivered as prescribed. Relative range verification allows intra-fraction monitoring of Bragg peak spacing to ensure full coverage with each fraction, as well as inter-fraction monitoring to ensure all fractions are delivered consistently. To validate the proposed filtered Interaction Vertex Imaging method for relative range verification, a $^{16}$O beam was used to deliver 12 Bragg peak positions in a 40 mm poly-(methyl methacrylate) phantom. Secondary particles produced in the phantom were monitored using position-sensitive silicon detectors. Events recorded on these detectors, along with a measurement of the treatment beam axis, were used to reconstruct the sites of origin of these secondary particles in the phantom. The distal edge of the depth distribution of these reconstructed points was determined with logistic fits, and the translation in depth required to minimize the $\chi^2$ statistic between these fits was used to compute the range shift between any two Bragg peak positions. In all cases, the range shift was determined with sub-millimeter precision, to a standard deviation of the mean of 220(10) μm. This result validates filtered Interaction Vertex Imaging as a reliable relative range verification method, which should be capable of monitoring each energy step in each fraction of a scanned heavy-ion treatment plan.

## 1 Introduction

While significant advances in cancer treatment have reduced patient mortality (Brenner *et al* 2020), cancer is still the leading cause of death in most developed countries, and the second-leading cause of death worldwide (Bray *et al* 2018, Roth *et al* 2018). Half of cancer patients undergo radiation therapy

(RT) as part of their treatment, often in combination with other modalities such as surgery or chemotherapy (Barton 2014, Tyldesley *et al* 2011). While traditional X-ray photon radiation therapy is still the most common type, heavy-ion therapy (HIT) is an established clinical method which uses charged particle beams such as $^4$He, $^{12}$C, or $^{16}$O to deliver the prescribed dose (Sokol *et al* 2017, Tsujii *et al* 2004, Amaldi and Braccini 2011, Scifoni *et al* 2013). An ion beam produces a characteristic dose distribution in a patient, with low entrance dose, high dose at the beam endpoint—also called the Bragg peak (BP)—and very little exit dose (Amaldi and Braccini 2011, Tessonnier *et al* 2017). HIT has several advantages over traditional RT, and has continually gained popularity over the past several decades (Amaldi and Braccini 2011, Amaldi and Kraft 2005). These advantages include the high precision with which the BP can be positioned on a tumour by simply changing the energy of the beam, and the increased effectiveness of ion beams at killing cancer cells (Amaldi and Kraft 2005, Krämer *et al* 2000, Laprie *et al* 2015). In contrast, the X-ray beams used in traditional RT deliver maximum dose near the surface, and an exponential decrease in dose at increased depth (Amaldi and Kraft 2005, Krämer *et al* 2000). Importantly, these X-ray beams do not stop in the patient, so a significant exit dose is present (Amaldi and Kraft 2005). The precision of HIT makes it superior to traditional RT in many cases, including in pediatric patients where the risk of secondary cancer from radiation damage to healthy tissue is greatest, and for treatment of tumours in or near sensitive organs, such as the head and neck (Tsujii *et al* 2004, Laprie *et al* 2015).

HIT is often administered via active (scanned) beam delivery (Amaldi and Braccini 2011). Here, a monoenergetic pencil beam is 'painted' over the tumour, through magnetic guidance or mechanical motion (Haberer *et al* 1993). The scanned beam covers the full lateral extent of a longitudinal slice in the target volume, using either a set of discrete BPs or a continuous pattern. To achieve full longitudinal coverage, multiple beam energies are used to adjust BP depth and treat all slices of the target sequentially. These scanned treatments are highly conformal, as each slice may use a different scan pattern to best follow the contour of the tumour at that depth.

To achieve maximum benefit from HIT, treatment plans and dose delivery must be similarly precise. While the physics of beam interactions in common materials are well-understood, the complex and dynamic internal structure of patients makes it challenging to determine the beam energy which places the BP precisely at the desired depth (Henriquet *et al* 2012, Gwosch *et al* 2013, Finck *et al* 2017, Jäkel *et al* 2001). To ensure that dose delivery matches the treatment plan as precisely as possible, and to monitor any deviations, robust range verification (RV) techniques are required (Henriquet *et al* 2012). Incorrect BP positioning is doubly problematic: the tumour is not treated as directed (leading to decreased tumour control) and healthy tissue is exposed to significant radiation damage (leading to increased incidence of side effects or secondary tumours) (Finck *et al* 2017).

Each BP energy in a scanned treatment is delivered consecutively, allowing independent monitoring of each slice, or possibly each individual BP position. Monitoring BP depth for each slice in a scanned treatment, or even difference in BP depths, could detect transient range errors appearing partway through treatment. Such errors might derive from patient or organ motion on the scale of minutes, or tissue inhomogeneities affecting beam range in some BP positions (Ammazzalorso *et al* 2014). Comparing the measured depth for two BPs targeted to the same position from different fractions can also achieve inter-fraction RV. This inter-fraction comparison may occur early in the delivery process, based on a single longitudinal slice. A key advantage of these relative RV techniques is that they depend only on the change in BP position between pencil beams to provide this valuable feedback, rather than requiring an absolute position, which is more difficult to measure precisely. Ideal HIT RV for either intra- or inter-fraction monitoring would provide real-time feedback on BP position, allowing treatments with incorrect positioning to be aborted before significant damage occurs. As even small errors in BP position may lead to significant negative patient outcomes, this ideal method must be sensitive to changes in position of one millimeter or less (Gwosch et al 2013).

Several HIT RV methods have been previously investigated, the majority of which are based on detection of radiation produced along the beam path or at the BP. The most mature of these methods are based on positron emission tomography (PET), with clinical implementation in some facilities (Parodi and Polf 2018). Beam interactions in the patient produce positron emitters, the distribution of which can be measured during or after treatment, with reaction products acting in lieu of the usual PET radiotracer (Amaldi and Braccini 2011, Enghardt et al 2004). However, the low detectable activity produced by $^{12}$C irradiation, relative to typical PET imaging using injected radiopharmaceuticals, places limits on the achievable sensitivity of the method. Additional complexities are introduced due to the inherent time delay of positron emission from the half-life of the produced positron emitter. Because of these delays, most PET implementations record data only after treatment is completed; these methods are appropriate only for verification of the entire fraction, or for inter-fraction comparisons (Handrack et al 2017). Recent studies using 'in-beam' PET allow limited collection of online data during irradiation, which would in principle allow intra-fraction comparison and relative RV. However, current in-beam implementations require additional data collection after treatment to produce accurate results (Pennazio et al 2018). An additional complication from time delay is washout: before positron emission occurs, biological activity may 'wash out' positron emitters from the region in which they were produced, reducing or diluting the measured signal. These washout effects can reduce PET RV from its ideal millimetric precision to worst-case values of 4-5 mm uncertainty (Handrack et al 2017). To perform absolute RV with PET, measured activity distributions are typically compared to those predicted by Monte Carlo simulations, including models for tissue activation and washout effects. As most PET RV implementations require either additional time in the treatment room, or transport to a dedicated PET facility for scanning, PET is not commonly performed for all fractions in a treatment plan (Parodi and Polf 2018, Handrack et al 2017). As such, there is a clear need for RV using prompt radiation, which can collect all required data during treatment, and allow for intra-fraction and more complete inter-fraction monitoring, all while avoiding the constraints on patient time and facility space imposed by PET.

Significant work has also been conducted with prompt gamma rays emitted at the instant of beam-patient interaction, both originating from tissue and from implanted tumour markers (Krimmer et al 2018, Magalhaes Martins et al 2020, Parodi and Polf 2018). While gamma rays easily exit the patient, they are comparatively difficult to detect, even more so for higher energy photons. This low detection efficiency, also a challenge in PET RV, is an inherent consequence of the uncharged nature of photons. For a photon to induce a signal in a detector, it must interact and locally ionize the material. Because these interactions are comparatively rare, photon detectors with reasonable efficiency must be large, or made from heavy high-Z materials. It is also very difficult to determine a photon's position of origin, as such measurements either require additional bulky collimators, which further limit detection efficiency in favour of admitting photons from a single direction only, or coincident detection of a scattered photon and later total absorption. This coincidence method, used in Compton cameras, follows the known kinematics of Compton scattering to reconstruct the initial trajectory of a photon that is first scattered, and later absorbed. However, the need for multiple low-probability detections again limits detection efficiency, and reduces the total number of photons which may be measured. In contrast, detection of prompt protons and other light charged particles, emitted as a result of the highly energetic beam interacting with patient tissue, can be easily detected. As these charged particles are directly ionizing, efficiencies can reach 100% in thin semiconductor detectors made from materials such as silicon. Because these detectors are thin, particles of intermediate or high energy do not typically stop in the detector, but often pass through with minimal deflection. If the same particle interacts with a second detector further from the patient, the trajectory between these two detection events can be extrapolated back into the patient. This method can be highly precise, with segmented detectors reporting charged particle positions with uncertainty on the order of 50 μm, and allowing sub-millimeter precision in both the reconstructed trajectory and in extrapolated quantities such as the lateral positioning of a treatment beam (Reinhart *et al* 2017, Félix-Bautista *et al* 2019).

Interaction Vertex Imaging (IVI) is a proposed HIT RV method which uses prompt secondary protons and other light ions (secondary particles) to reconstruct the positions of nuclear reactions (the 'interaction vertices') within a patient. These protons are mostly emitted after fragmentation reactions of the beam and tissue along the trajectory of the beam. In an ideal implementation, the reconstruction of interaction vertices could take place in real time, during treatment, to provide immediate feedback on BP position. IVI was first proposed in 2010 (Amaldi *et al* 2010), and a number of feasibility studies have been conducted for lateral monitoring of the treatment beam (Gwosch *et al* 2013, Reinhart *et al* 2017, Félix-Bautista *et al* 2019) and range verification (Henriquet *et al* 2012, Gwosch *et al* 2013, Finck *et al* 2017). Recently, IVI techniques have been integrated into clinical trials for $^{12}$C treatment, for both range (Fischetti *et al* 2020) and lateral monitoring (Toppi *et al* 2021). To date, RV methods have consistently reported maximum precision of 1-2 mm, with sub-millimeter precision of 0.2-0.8 mm achieved for lateral monitoring only. However, these previous studies have focused on absolute range verification, which limits precision due to inherent systematic uncertainties.

Previous studies of IVI have also primarily used small and highly-segmented detectors (Gwosch *et al* 2013, Finck *et al* 2017), which provide excellent spatial resolution for the detection of secondary particles, at the expense of high electronic complexity. When detectors using small pixels are expanded to larger sensitive areas (Henriquet *et al* 2012), the number of pixels, and thus the number of readout channels, increases rapidly, placing greater demands on the processing and readout systems which make real-time RV more difficult to implement. However, larger sensitive areas are valuable for clinical applications, where the number of primary ions is determined by the treatment plan, and an RV method must collect sufficient data to determine BP range with the limited number of secondary particles produced by these primaries. Larger sensitive areas will allow the collection of more secondary particles in an otherwise-identical treatment.

As recent studies using scintillator detectors have indicated that the spatial resolution provided by small pixels may not be required in clinical settings (Fischetti *et al* 2020, Toppi *et al* 2021), an opportunity exists to reduce the spatial resolution in silicon detectors, to allow increased sensitive areas while maintaining or even reducing readout complexity. Reduced readout complexity for large sensitive area silicon detectors is available in strip-segmented designs, where the number of readout channels scales with perimeter, rather than area; and with position-sensitive detectors, where the entire sensitive area is monitored using only four readout channels, one per lateral edge (Soisson *et al* 2010). One goal of this study is to investigate whether these lower-cost and reduced-complexity systems are capable of achieving sufficient RV performance for real-time IVI.

Recent work by the authors has proposed a novel implementation of IVI which achieves sub-millimeter precision in the measurement of range differences between two BP positions, using data generated through Monte Carlo simulation (Hymers and Mücher 2019). This filtered IVI (fIVI) method selects only secondary particles which can be reconstructed with high precision, and are most likely to have originated from an interaction of the primary treatment beam with the patient. The current study was designed to allow a first experimental investigation of the fIVI algorithm in a simple homogeneous phantom, using low-complexity position-sensitive detectors, and validate this technique for reliable relative sub-millimeter determination of BP depth differences, in the context of intra-fraction monitoring.

# 2 MATERIALS & METHODS

## 2.1 OVERVIEW

To validate fIVI for reliable relative sub-millimeter RV, a new detection setup was created at the National Superconducting Cyclotron Laboratory (NSCL), Michigan State University, Michigan, USA. A $^{16}$O beam with an initial nominal energy of 150 MeV u$^{-1}$ was degraded to different final energies to model shifts in BP position. The degraded beam then entered a homogeneous poly-(methyl methacrylate) (PMMA) phantom. Outgoing charged particles were detected using a newly developed two-arm silicon tracker using position-sensitive silicon detectors (PSDs). Using the technique previously described by Hymers and Mücher (2019), the BP depth differences between all beam combinations were reconstructed, and the overall performance evaluated.

## 2.2 SIMULATION DETAIL

Monte Carlo simulations were used prior to experiment to investigate options for detector positioning, as well as during analysis to provide a rough energy calibration. This calibration is used for illustrative purposes only, as the fIVI analysis procedure can be performed in arbitrary energy units. The simulation was the same as described by Hymers and Mücher (2019), based on Geant4 version 10.02 (Agostinelli *et al* 2003). Information on physical processes was provided by the pre-packaged physics list QGSP_BIC, which provides appropriate results for beams below 200 MeV u$^{-1}$ using a binary cascade model (Finck *et al* 2017). While this model is known to not perfectly reproduce the experimentally-recorded number of secondary particles, it has been previously shown to produce a vertex distribution shape that matches experimental data (Henriquet *et al* 2012, Finck *et al* 2017).

## 2.3 EXPERIMENTAL FACILITY

Data collection took place at the Single Event Effects Test Facility (SEETF) in NSCL. NSCL is capable of generating primary beams from $^{16}$O to $^{238}$U, with energies as high as 170 MeV u$^{-1}$, although the maximum energy varies for each isotope. Ions were accelerated in a pair of coupled superconducting cyclotrons: initial acceleration was performed by the K500 cyclotron, after which the ions were injected into the K1200 cyclotron for acceleration to the final energy (Gade and Sherrill 2016). In the present experiment, a $^{16}$O beam was accelerated to 149.41(10) MeV u$^{-1}$. Beam current was controlled by an attenuator between the ion source and the K500 cyclotron, which served to reduce the impact of attenuation on beam homogeneity at the phantom (Ladbury *et al* 2004). The fully accelerated beam was guided through the A1900 superconducting fragment separator and delivered undegraded to the SEETF beamline. Beam energy and intensity measurements were made between irradiations using a Faraday cup inserted into the SEETF beamline upstream of the multi-purpose user setup.

To modulate beam energy, a range shifter device integrated into SEETF was used. This device allowed either of two different absorber plates to be inserted, and rotation of these plates to adjust their effective thickness. For this experiment, two aluminum degraders were installed, with thicknesses of 1.8 mm and 2.6 mm. Degrader insertion and rotation was controlled remotely using stepper motors with a precision of $\pm 0.5°$. These motors were fully disabled during data collection to reduce electromagnetic interference in the data acquisition system. The degraded beam exited the SEETF vacuum window, a 75 µm zirconium foil, and entered the experimental setup, in air. An overview of the user area setup in SEETF, including the degraders installed in the evacuated beamline, is shown in Figure 1.

## 2.4 BEAM SELECTION

A $^{16}$O beam with a nominal energy of 150 MeV u$^{-1}$ was selected for primary irradiation. While $^{16}$O is not currently used for clinical irradiation, it shares many of the same benefits as the clinically-common $^{12}$C, and is of particular interest in the treatment of radioresistant hypoxic tumours (Tessonnier *et al* 2017). Treatment planning with $^{16}$O is an area of active research interest (Sokol *et al* 2017). At the same time, the heavier $^{16}$O ions are of interest for IVI applications because of their greater total beam energy, transferring more energy to a single proton produced by a fragmentation reaction. The greater production of high-energy secondary particles potentially increases the number of particles which escape the phantom and reach the detectors, while reducing the overall energy straggling.

As the feature of interest was the behaviour of the vertex distribution with changes in BP position, the subclinical energy of 150 MeV u$^{-1}$ was selected to allow focus on this region only. Pre-experiment simulations indicated that the distal edge shape in the BP region was similar at higher energies, with some variance in slope due to the increased energy spread associated with longer travel distance in PMMA. While this configuration differs significantly from a clinical setup, it was deemed appropriate to validate the fIVI implementation and associated range verification method in this simplified setting for $^{16}$O beams before continuing to more clinical configurations.

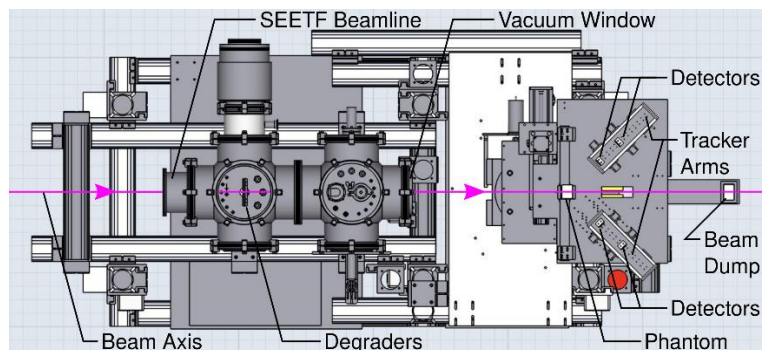

Figure 1: Setup in experimental hall, with critical elements of apparatus labelled. To the left, additional upstream elements of the SEETF beamline exist, but are not shown in this drawing. One grid square is equal to five centimeters. Detail view of detector and phantom arrangement, as mounted to the plate on the right of the drawing, are shown in Figure 2.

Online measurement of the beam indicated that the delivered energy was 149.41(10) MeV u$^{-1}$, with a beamspot size of 2.4 mm FWHM (full width at half maximum). Beam current varied between 20 pA and 50 pA (1.57×10$^7$ to 3.90×10$^7$ ion s$^{-1}$) during data collection. This intensity, an order of magnitude lower than typical clinical irradiation, was chosen to limit the count rates on the PSDs in this initial trial.

## 2.5 DETECTION SYSTEM

For this setup, the MS PSD DP04 2E/2E position-sensitive silicon detector, manufactured by Micron Semicondutors, was selected. These detectors provided a 20 mm × 20 mm sensitive area of 300(15) μm thickness, in a low-cost and low-complexity detection system. Although the 300 μm thickness is larger than ideal for single-particle tracking (Hymers and Mücher 2019), the impact of in-detector scattering on overall range verification performance was deemed acceptable based on simulation results.

Rather than being divided into segments to provide information on the position of an interaction based on the segment which is triggered, PSDs include a resistive layer. This layer divides the energy deposit from a given interaction between opposite edges of the detector in proportion to the position of the interaction, such that interactions closer to a given edge will produce a greater signal on that edge's readout. However, this layer also introduces a long charge collection time, which reduces the count rate that these detectors can maintain. To provide a two-dimensional readout, both the horizontal and vertical axes of the detector use position-sensitive detection, allowing readout of the entire sensitive area with only four channels (one channel per 20 mm lateral edge), a significant reduction in electronic complexity as compared to the more common pixelated detectors. To record an interaction, a four-channel coincidence is required, which allows the position of the interaction to be reconstructed as the weighted average of the signals at all four edges. Under ideal conditions, these detectors are capable of 200 μm position resolution (Soisson *et al* 2010); however this result is highly sensitive to noise, as the position is reconstructed from the four individual edge readouts.

Each detector channel was connected to one channel of a Mesytec MPR-16L preamplifier, which converted the single-conductor readout to a differential output. These preamplifiers also provided power to the PSDs, applying a bias of 40.0(1) V across the thickness of the detector. Each preamplifier channel was connected to one channel of an eight-channel CAEN desktop digitizer; both models DT5725 and DT5730S were used. These digitizers provided fast (order nanosecond) timing resolution, and acquisition through CAEN's CoMPASS frontend, which produced output in ROOT format for later analysis (Brun and Rademakers 1997).

To facilitate the two particle detections required by the fIVI algorithm, two light-tight aluminum tracker boxes were designed and built, with mounting positions for multiple detectors. Two detectors were mounted in each of these boxes, in parallel planes 120 mm apart. Each box with its two detectors comprised one arm of the overall two-arm design. As each tracker contained only two detectors, for a total of eight channels, each tracker was connected to a single digitizer only. The tracker walls had a thickness of 6.4 mm, with the exception of the front panel (facing the phantom) which was replaced with a 16 μm aluminum foil. This thin foil minimized opportunities for deflection of secondary particles when entering the tracker, while still protecting the detectors from light in the experimental hall.

## 2.6 SETUP AND ALIGNMENT

Trackers were positioned at a 45 degree angle from the primary beam axis, as shown in Figure 2 a). The PSDs' reduced electronic complexity comes at the cost of reduced count rate capabilities, due to the entire sensitive area being involved in acquisition of each hit. The 45 degree angle was selected based on Monte Carlo simulations to reduce the secondary particle flux, and consequently the incidence of multiple detection events which would prevent the accurate reconstruction of hit position. Larger off-axis angles also provided less longitudinal uncertainty, for geometric reasons. Angles larger than 45 degrees were undesirable, due to a further reduction of secondary particle flux leading to fewer candidates for secondary particle reconstruction.

As the number of secondary particles escaping the phantom at 45 degrees is an order of magnitude smaller than the number escaping at 10 degrees (Finck *et al* 2017), some compensation is necessary. As the PSDs used in this subclinical study are of similar total size to previous studies, but expect to experience lower count rates, it was necessary to collect data for a longer duration. To ensure that results from this study were statistically comparable to previous work, an upper limit on reconstructed interaction vertices was set at $10^3$ vertices $mm^{-1}$ in the longitudinal direction, a value comparable to previous studies (Henriquet *et al* 2012, Finck *et al* 2017). In a clinical setup, a larger sensitive area could perform the same compensation without requiring an increase in irradiation time or total number of primary ions delivered to the patient, discussed further in section 4.2.

Fixed tracker positioning relative to the phantom was achieved by mounting both trackers and the phantom to a single baseplate, with mounting holes drilled by a computer-controlled system. While each hole was drilled with precision on the order of 100 μm, the assembly of each tracker from multiple pieces led to overall uncertainty in setup alignment of order 1 mm. The entire system of phantom and detectors was placed in air. No additional alignment was performed, as it was presumed that the detectors and phantom remained fixed during all measurements, a sufficient condition for relative RV which requires no absolute reference point for depth.

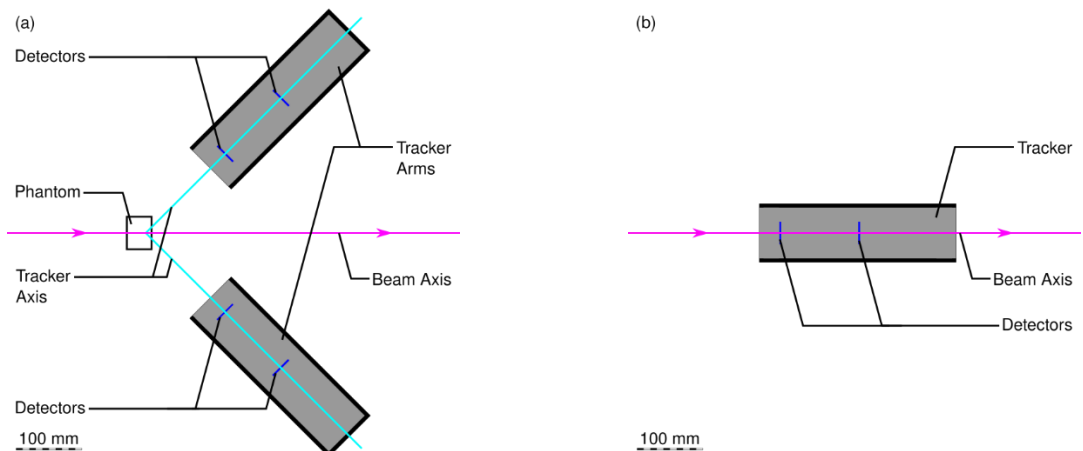

Figure 2: a) Scale schematic of experimental setup for fIVI measurement. The beam (magenta) entered the PMMA phantom (black outline) from the left of the figure. Each tracker (grey) contained two detectors (blue), and had its central axis (cyan) aimed at a 30 mm depth in the phantom. b) Scale schematic of experimental setup for beam axis measurement. The PMMA phantom was removed, and the rear panel of the tracker was replaced with a foil window, so the full-energy beam could continue through the tracker to the beam dump (not pictured). For context on positioning of this arrangement relative to the full SEETF setup, see Figure 1.

The 52 mm × 52 mm, 40 mm thick PMMA phantom was held in place from the bottom by a clamp, to prevent any interference of the mounting system with beam delivery. Each tracker was aligned such that its central axis, passing through the center of both mounted detectors, intersected with the nominal beam axis at a depth of 30 mm in the phantom. Alignment of the beam axis was verified using the optical alignment system integrated into SEETF. A direct beam axis measurement was also performed at the end of the data collection period. For this measurement, one tracker was moved into the beam axis path, with alignment again verified by the SEETF optical alignment system. For this beam axis measurement only, the phantom was removed and the rear panel of the detector was replaced by a 16 µm aluminum foil, to allow the undegraded beam to continue through the detector into the beam dump. The modified setup for this beam axis measurement is illustrated in Figure 2 b).

## 2.7 DATA COLLECTION

Data were collected at each of twelve BP positions, over a total depth of 7.4 mm. The range of each BP was determined using LISE++ (Tarasov and Bazin 2016), and verified using SRIM (Ziegler 2004). In LISE++, developed at NSCL to model particle yields in multi-stage beamlines, the irradiation apparatus was modelled. This model included the degrader installed in the SEETF vacuum chamber, the zirconium exit window, a path length of 60 cm in air, and the PMMA phantom. Each planned degrader setting was evaluated in LISE++ to determine the resultant energy distribution incident on the phantom. The mean beam energy reported by LISE++ was recorded, and the associated range evaluated in the LISE++ physical calculator and in SRIM, to determine the position of the beam endpoint. In all cases, both calculations produced similar depths in the PMMA phantom, with nominal range differences of less than 50 microns. Uncertainty in BP position was derived from longitudinal straggling of a monoenergetic beam, and from the broadened distribution incident on the phantom after the degrader. Information on the various configurations studied is shown in Figure 3.

Data were collected in each position for one run of at least 300 seconds; some positions also included one or more shorter irradiations of approximately 100 seconds. This extended duration, as compared to a clinical irradiation, was found to provide appropriate scaling to account for the low efficiency and reduced count rate of the PSDs, as discussed in section 2.6.

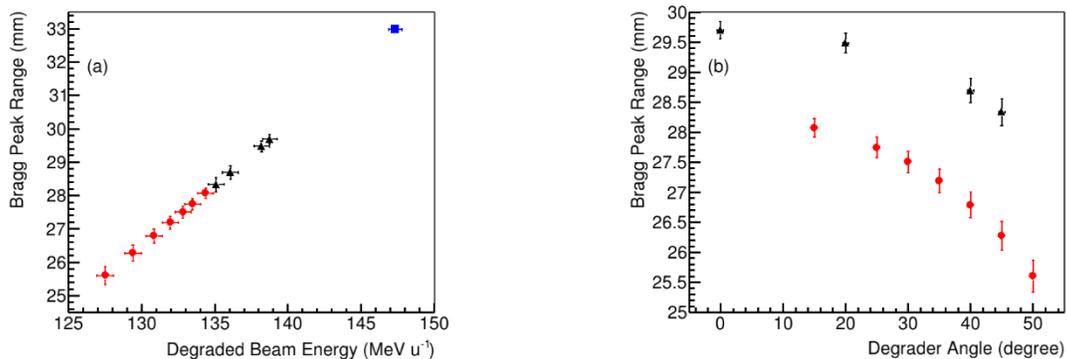

Figure 3: a) Degraded beam energies incident on PMMA phantom, as computed by LISE++ based on an undegraded energy of 149.41 MeV u$^{-1}$, and associated Bragg peak range. b) Degrader angle and associated Bragg peak range, using the same computation. For both panels, energies and ranges achieved using the 1.8 mm degrader are shown with black triangles, and the 2.6 mm degrader with red circles. The undegraded beam in a) is shown with a blue square.

A typical irradiation process began with the beam stopped in the SEETF Faraday cup, prior to encountering the degraders or exiting the vacuum window. The motors controlling degrader position were activated, and the degrader was adjusted to the prescribed angle. Once the degraders were adjusted, the motors were disabled. Data acquisition was started manually as the SEETF shutter was opened and Faraday cup removed. Typical count rates per channel for front detectors in both trackers were 1.5 to 5.0 kHz, while the rear detectors exhibited a lower rate of 0.5 to 2.0 kHz per channel. After completing the run, data collection was stopped manually, the shutter was closed, and the Faraday cup reinserted. If beam current had dropped below 20 pA, the time between runs was used to tune the beam to higher intensity, up to 50 pA, as intensity tended to decrease over time. The same detector configuration and PMMA phantom was used for all runs. Data collection was completed from deepest to shallowest BP depth studied.

For the beam axis measurement, the beam intensity was reduced below the SEETF measurable threshold, to prevent damage to the PSDs and allow sufficient time for charge collection and acquisition of discrete events in these low-rate detectors. The irradiation procedure was similar to primary fIVI data collection. This measurement was completed for the undegraded beam only.

## 2.8 ANALYSIS AND RECONSTRUCTION

Analysis was performed offline, following data collection. Raw output from each PSD was converted into hits with a two-dimensional position in the plane of the detector, using the typical weighted average of the four edge readouts. Multiple detection events for which the total energy measured on the front and rear differed by more than 200 keV were rejected. Each hit was assigned a timestamp equal to the earliest channel in the hit to trigger. Only events which triggered all four channels of the PSD exactly once in a 300 ns coincidence window were considered for reconstruction. This 300 ns window was chosen to admit as many events as possible given the long charge collection time of the PSDs, while not being so long as to conflate events caused by two different secondary particles, where more than one trigger event occurred on a single channel within the coincidence window. These two-dimensional hits were then converted to three-dimensional positions in the lab frame, using the known positions of the detector relative to the phantom. These detector hits were used as input to the fIVI algorithm described below. The beam axis measurement produced two points in the lab frame; the vector passing through both of these points was used as the beam axis parameter in reconstruction.

The fIVI method first described by Hymers and Mücher (2019) was used to convert information on secondary particles into a longitudinal vertex distribution describing the depth of the Bragg peak in a target. This method is similar to the reconstruction used by Finck *et al* (2017), with the addition of energy and spatial filters to preferentially reject scattered particles. While the previous description of the fIVI method included cuts on both the total kinetic energy and the deposited energy of each secondary particle, satisfactory performance was achieved in this case with a cut on deposited energy only, with events depositing less than 200 keV on a single detector being rejected. As in the work by Finck *et al*, a secondary particle interacting with two detectors within a software coincidence window (300 ns in this work) formed a track, defined as the vector between the positions of the two hits. All possible tracks were generated based on the set of coincident hits. These tracks were then extended toward the central axis of the treatment beam. For each track, its point of closest approach to the beam axis was computed, as well as the corresponding closest point on the beam axis itself. Another filter not present in Finck *et al* was used to reject tracks for which these points were separated in three dimensions by more than twice the beamspot width (i.e. 6 mm separation for a 3 mm FWHM beam). The remaining tracks estimated the site of secondary particle production (the 'interaction vertex'), as in Finck *et al*, to be the midpoint between these two points of closest approach. For detectors with a large sensitive area, and particularly at high event rates, multiple secondary particles may interact with the detectors in coincidence, allowing for false tracks to be formed from interactions of two different secondary particles, in addition to the true tracks formed by each particle. Rejection of false tracks was achieved through the criterion on closest approach to the beamline, which also rejected tracks corresponding to secondary particles which scattered in the phantom before reaching the detectors.

As the experimental PSDs were much smaller than the detectors in the previous simulation-based study, the probability of two-particle coincidence events was found during simulation to be extremely low, limiting the utility of triangulation IVI, a method which refines the interaction vertex position using two coincident secondary particle tracks produced from the same fragmentation reaction. Experimentally, only single-particle reconstruction was used, and no candidate events for triangulation IVI were investigated.

The longitudinal distribution of these interaction vertices (the 'vertex distribution') was used to compare two BP depths; making this comparison with high accuracy and precision was the goal of this investigation. The distal edge of the vertex distribution was defined as the region where the vertex count in a bin decreased to zero from the local maximum at greatest depth. For this identification, a local maximum was defined as a bin containing *n* entries, separated from any bin containing $m > n$ entries by at least one bin containing $l < n - 100$ entries (the 'edge bin'). The value 100 was chosen as the square root of $10^4$, the approximate number of entries in a typical vertex distribution. When comparing two distributions, one was designated the 'reference' measurement, and the other the 'test' measurement. A logistic function (Equation 1) was fit to the distal edge of each measurement. In this function, the parameter *h* defines the height of the curve, by which the upper and lower asymptotes are separated, and the parameter $y_0$ defines the vertical offset, the height of the lower asymptote. The

$$f(x) = y_0 + \frac{h}{1 + \exp(-w(x - x_0))} \qquad (1)$$

width parameter *w* controls the width of the curved region between asymptotes, to match the shape of the distal edge, while the horizontal offset $x_0$ sets the inflection point of this curved region. The fit region was from the edge bin to a position downstream of the distal edge, where no further interaction vertices are reconstructed. This method enforces a well-defined lower asymptote in the fit region, which contributes to the consistency of the fit shape in the distal edge region. For all BP positions, this downstream edge was beyond the phantom, at a depth of 10 cm, and the histogram of interaction vertices had a 1.0 mm bin size. This fit method was consistent with the previous study (Hymers and Mücher 2019). To allow comparisons between data sets of different sizes, such as a comparison between a 100 s and a 300 s irradiation, and to reduce the effect of statistical error, the fit function for the test measurement was scaled and translated vertically to match the reference measurement. This scaling was performed such that both fit functions had equal values for the height (*h*) and vertical offset ($y_0$) parameters, and differed only in the horizontal parameters of offset ($x_0$) and width (*w*). The scaled fit to the test measurement was then horizontally translated in 100 µm increments, to identify the shift which best agreed with the fit function to the reference measurement. For each position, a discrete $\chi^2$ statistic, sampling every 100 µm, was calculated using the reference fit as the expected value and the translated test fit as the trial value. The translation which produced the best agreement, as signified by the lowest value of the $\chi^2$ statistic, was defined as the reconstructed depth difference between the two measurements, or the range shift of the test measurement relative to the reference measurement. Comparisons between these entire fits, rather than a comparison of the longitudinal position $x_0$ only, provides a more accurate measurement of the depth difference by reducing the impact of position-dependent differences in the exit path length for secondary particles which distort the shape of the distal edge, contributing to differences in the parameter *w*.

To estimate an energy calibration using simulated data, and for comparison of secondary particle hits between simulation and experiment, the simulation geometry was configured to match the experimental setup. Rather than modelling the entire beamline and degrader apparatus, the energy of primary particles was set based on mean energies and widths calculated by LISE++ for particles exiting the beamline vacuum window at each degrader position. Simulations were completed for $5 \times 10^8$ primary ions, which provided similar numbers of detected secondary particles to experiment. To simulate the effect of false coincidence events between secondaries produced by different primary particles, a global time offset was applied to each hit on a simulated detector. For comparisons between simulation and experiment, this offset was selected to correspond to a beam current of 50 pA, matching the maximum experimental intensity. Results were output in ROOT format, and analyzed using the same process as experimental data. To model the precision and nonuniform response observed in the PSDs, a radial efficiency correction was applied to simulated data before reconstruction, based on the observed detector response, shown in Figure 7.

## 3 RESULTS

### 3.1 BEAM MONITORING

Before fIVI reconstruction can be performed, the average position and width of the treatment beam must be measured, for use as reconstruction parameters. Figure 4 shows the energy deposition spectra for the front and rear silicon detectors during the axis measurement of the undegraded beam. As the preamplifier sensitivity was changed for this measurement only, to account for the higher energy deposit from the primary beam, no energy calibration is available for these spectra, which are instead presented in the uncalibrated units of the data acquisition system's least significant bit (LSB units). In the raw spectra, multiple peaks are evident, indicating the possible presence of multiple different particle energies or charges. However, only the dominant peak, with the highest deposited energy, is correlated between the front and rear detectors of the tracker; this peak corresponds to the tracked primary beam. The similar energy of this peak on both detectors indicates that energy loss is constant, as expected for a high-energy beam. A reduction in efficiency of approximately 50% is observed for the rear detector in Figure 4 b), as compared to the front detector in Figure 4 a). This reduction is primarily attributed to the mean beam position for the rear detector, shown in Figure 5, being further from the detector origin, which reduces efficiency as shown in Figure 7. The lower-energy peaks are uncorrelated between the two detectors of the tracker, and are believed to be an artifact of the noisy detection system, rather than physical particles; future studies will investigate these uncorrelated events in more detail. Applying the same track-formation algorithm as used in fIVI reconstruction, with a reduced 18 ns coincidence window between the two detectors, allows tracking of the primary beam path while neglecting spurious events. This coincidence cut alone eliminates the majority of the low-energy events shown in Figure 4 a) and b), and shrinks the measured beamspot in Figure 5 a) and b) to approximately the 3 mm nominal width of the experimental beam, in line with the reported value from the accelerator team for the beamspot size closer to the ion source. This cut also reduces the number of events in the primary peak, due to the poor efficiency of PSDs reducing the rate of observable coincidence events.

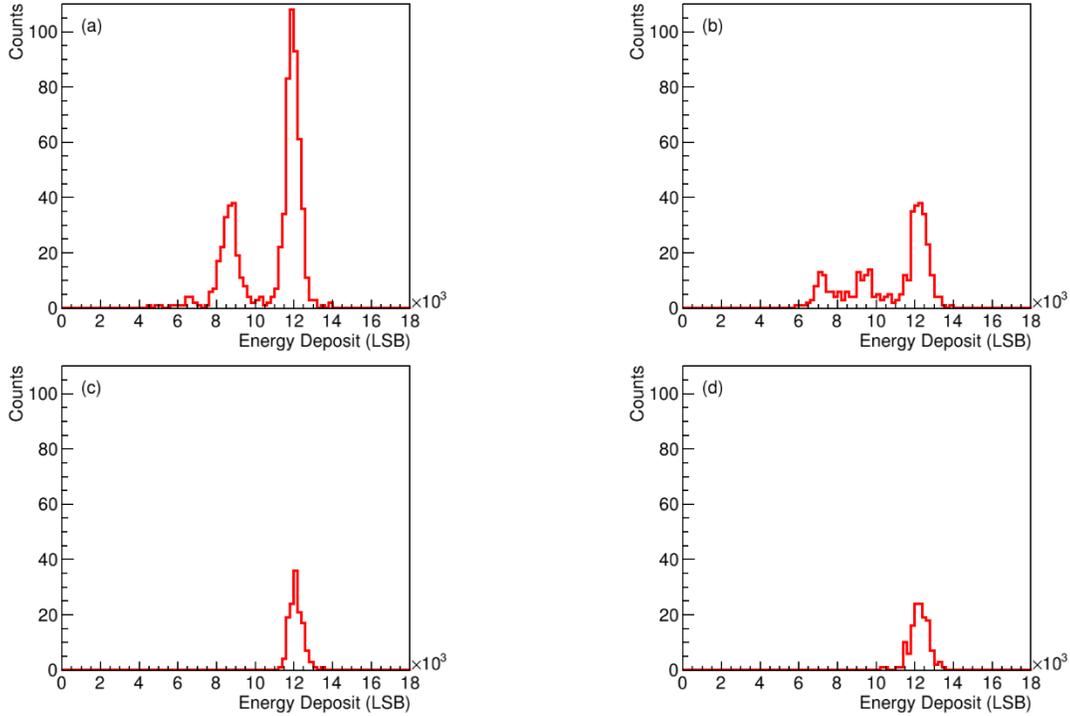

Figure 4: Energy deposition spectra for the beam axis measurement. a) Unfiltered spectrum for front detector. b) Unfiltered spectrum for rear detector. c) Filtered spectrum for front detector. d) Filtered spectrum for rear detector. Energies are presented in the uncalibrated units of the data acquisition system's least significant bit (LSB units).

To further clean up the position plot, an energy cut at 10 000 LSB units was applied to all tracks passing the coincidence cut, representing events for which the beam passed through both detectors without deflection or fragmentation. The result of this energy cut at 10 000 LSB units alone is similar to the result of the coincidence cut, emphasizing that these lower-energy peaks do not derive from direct interactions of the beam passing through the detector. The results of applying both cuts together are shown in Figures 4 and 5, c) and d); the similar distributions resulting from either cut applied alone are not shown.

Although a shorter coincidence time was used in the analysis of the beamspot size and position, as compared to the fIVI reconstruction, extending this coincidence window to the same 300 ns did not significantly affect the reconstructed beam position. The mean positions on each detector, as well as the 3 mm nominal beamspot width, were used as parameters in the fIVI reconstruction. The two mean points were used to define a beam axis trajectory in the same fashion as for secondary particles, while the beam width was used to define the allowable deviation of an interaction vertex from this axis. The mean positions computed for the beam axis are consistent with visual alignment performed during experiment setup using the SEETF laser alignment system.

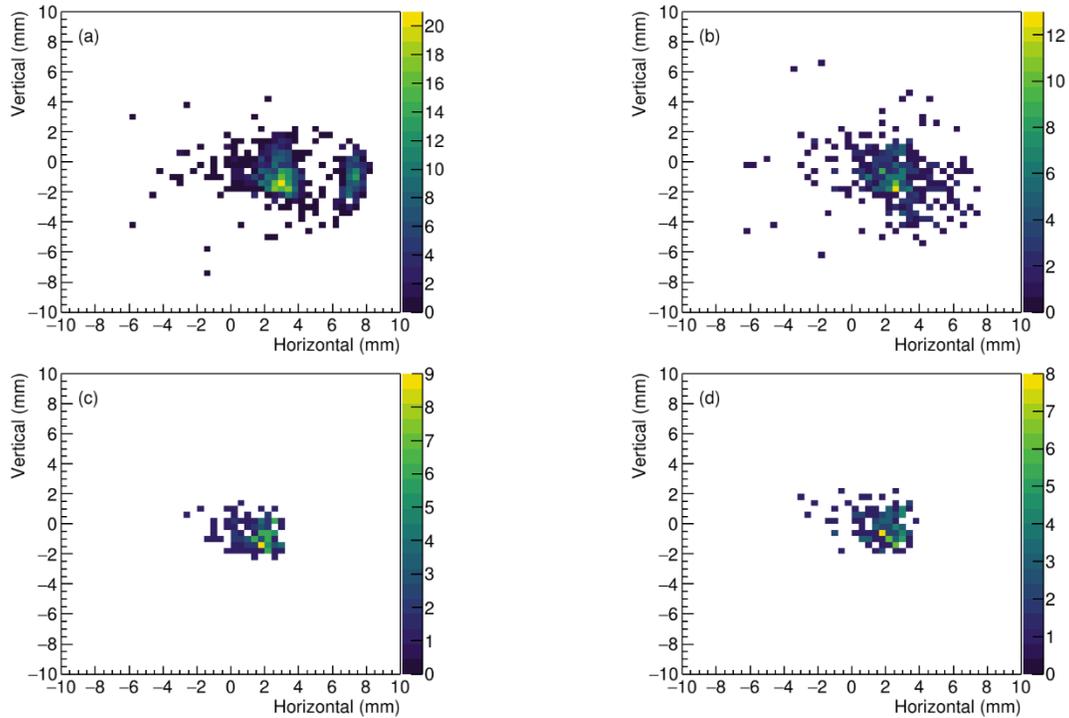

Figure 5: Position of beam interactions in detector plane, as viewed from the beamline. a) Unfiltered spectrum for front detector. b) Unfiltered spectrum for rear detector. c) Filtered spectrum for front detector. d) Filtered spectrum for rear detector. The cuts and filtration from Figure 4 shrink the beamspot to a measured size of 2.4 mm, close to the nominal width of 3 mm. The front detector (c) reported a mean horizontal position of 1.52 mm, and a mean vertical position of -0.71 mm, while the rear detector (d) reported a mean horizontal position of 1.82 mm, and a mean vertical position of -0.16 mm.

## 3.2 RAW FIVI SPECTRA

The energy spectrum for those individual secondary particle detection events which are candidates for fIVI reconstruction is shown in Figure 6. Each experimental event corresponds to coincident detection of exactly one trigger event on all four channels of a PSD placed at a 45 degree angle from the treatment beam axis. The total energy deposit is computed as the average energy deposit recorded by each pair of edges (top and bottom, and left and right). The agreement in shape between the simulated and experimental spectra was used to perform a rough energy calibration, which is presented for illustrative purposes; the fIVI analysis itself can be performed independently of the energy calibration. The energy cut in this work occurs to remove the low-energy tail below 200 keV, which occurs at an energy clearly evident in the uncalibrated spectrum. This spectrum is peaked around 500 keV, corresponding primarily to high-energy secondary particles experiencing linear energy loss through the silicon PSDs. This behaviour was similar for all four detectors at all BP positions.

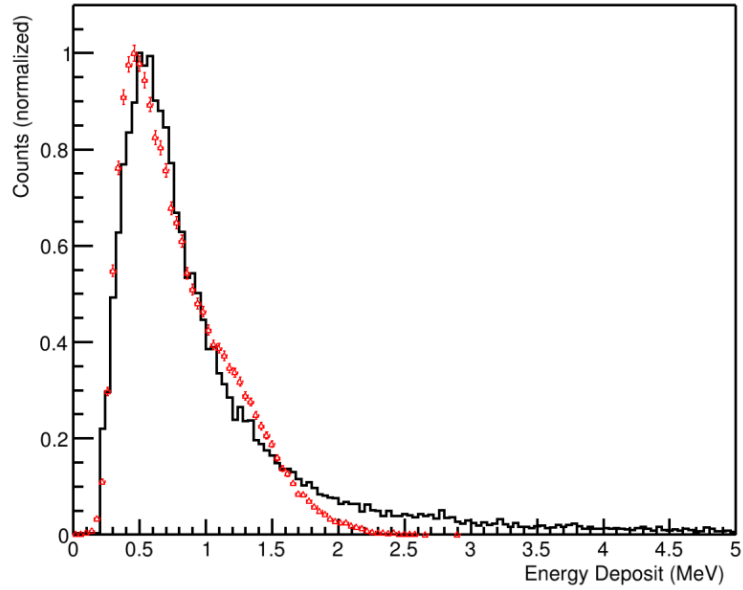

Figure 6: Energy deposit spectrum for secondary particle hits on a single front detector, from one arm of the tracker (red triangles), as compared to a simulated spectrum (black line). The spectra are normalized such that the most common incident energy has an intensity of one, to account for efficiency differences between simulation and experiment. The experimental data is for a BP depth of 28.3 mm, and is representative of data from all detectors at all BP positions. The significantly greater incidence of high-energy detection events (with an energy deposit above 2 MeV) in the simulated spectrum at similar depth is attributed to the simulation overestimating the number of heavier secondary ions (including deuterons, tritons, and alpha particles) which reach the detectors.

The count rate was higher by a factor of three for the front detector in each tracker than for the rear detector, as is expected from the larger solid angle covered by the front detector. This difference in count rates corresponds to the difference in trigger rates between front and rear detectors observed during data collection. Simulation data shows similar behaviour, although the discrepancy between front and rear counts is slightly larger than in experiment.

When the 2D reconstruction of hits from all four-channel coincidence events was completed, a significant difference was observed between the experimental results for hit position, shown in Figure 7 a), and the expected results for a segmented detector with 100% efficiency, shown in Figure 7 b). While the idealized detector shows a relatively uniform distribution of hits across the entire detector surface, with a small bias towards smaller off-axis angles, where flux of secondary particles is greater (Finck *et al* 2017), the experimental distribution of hits on the PSD shows a significant bias towards the center of the detector, with significant reductions in efficiency towards the edges and corners of the detector. This behaviour is believed to be due to the propagation time in the PSD's resistive layer being sufficiently long that events near to one edge are less likely to have both edges trigger within the 300 ns coincidence window, even with the modest beam intensity and large off-axis angle of 45 degrees. As propagation time differences are increased for interactions closer to the edge of the detector, the efficiency is further reduced for interactions closer to the edge. This phenomenon is compounded for events near the corners of the detector, where large differences in propagation time exist on both axes, further decreasing the likelihood of all four edges triggering within the same 300 ns coincidence window.

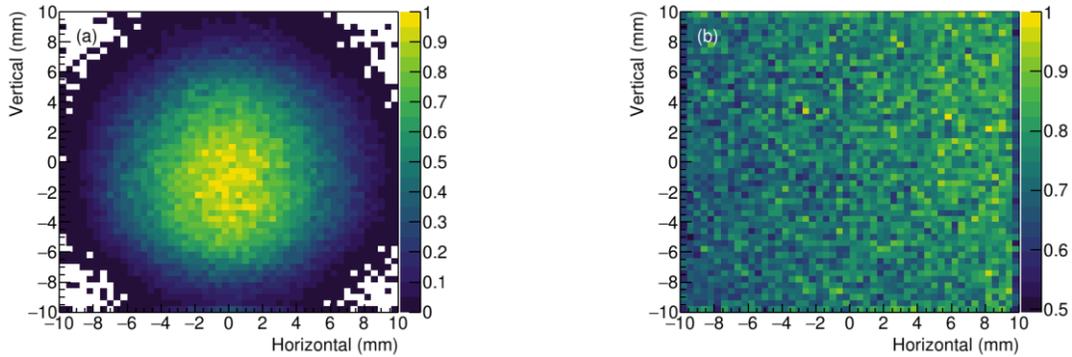

Figure 7: a) Experimental position of secondary particle interactions in detector plane for front PSD of left tracker, as viewed from the beamline side of the phantom. b) The same detector position, for a simulated segmented detector. In both panels, the origin is taken to be the center of the detector. The higher count rate to the right of the figure is expected, as secondary particles are more likely to be produced at smaller off-axis angles. Both panels are for a BP depth of 28.3 mm, using the same data shown in Figure 6.

The effect of this efficiency reduction on the range verification result is expected to be minimal, as interaction vertices in the region of the Bragg peak and the distal edge of the vertex distribution are those produced near the center of the field of view, which are the least affected by the efficiency correction. The fIVI method itself is independent of the shape of the detector, and uncertainty in the position of each hit is within acceptable bounds to produce a sufficiently accurate vertex distribution. However, PSD efficiency is relevant for future clinical applications, where the number of primary ions is limited by the treatment plan and cannot be increased to compensate for poor detector performance. The consequences of this result are discussed in section 4.2.

The impact of this radial efficiency dependence on overall efficiency of vertex reconstruction was estimated through two methods. In the first method, the rate of four-channel coincidences forming hit positions was compared to the per-channel rate of trigger events. A beam-off measurement of noise estimated the background trigger rate to be approximately 1 Hz, so all trigger events were assumed to derive from true secondary particle interactions. Comparing the number of trigger events per channel to the number of reconstructed hit positions via four-channel coincidence events yielded 20(1)% efficiency on the DT5725, and 34(2)% efficiency on the DT5730S. If the probabilities of detecting a secondary particle on the front detector and rear detector were independent, this would lead to efficiency of 4-12% for track formation, which requires two interactions on different detectors of the same tracker.

A second evaluation of efficiency, based on reconstruction of simulated data from a segmented detector, was performed, both with and without the radial efficiency correction suggested by Figure 7 a). This assessment yielded a vertex reconstruction efficiency of 12.8% for the PSD-corrected data, consistent with tracks originating near the center of the tracker's field of view passing through the high-efficiency center of both detectors, while tracks originating from other regions experienced an efficiency similar to the independent detection case.

### 3.3 VERTEX DISTRIBUTION

Vertex distributions are the longitudinal distributions of reconstructed interaction vertices along the nominal beam axis from Figure 1 and Figure 2, produced as described in section 2.8. This reconstruction used the beam parameters from section 3.1, along with secondary particle detection events from section 3.2. The nominal beam axis does differ slightly from the measured position shown in Figure 5; however, the resultant differences in Bragg peak depth along the longitudinal axis are negligible. Each tracker independently reproduced vertex distributions with the same shape. The presented vertex distributions combine data from both trackers to produce one overall vertex distribution containing all reconstructed interaction vertices. These distributions are composed of a comparable number of interaction vertices to previous studies (Gwosch *et al* 2013, Finck *et al* 2017), with an average of $1.4 \times 10^4$ reconstructed tracks used in each distribution. The largest data set, at the greatest BP depth of 33 mm, included $1.9 \times 10^4$ reconstructed tracks, while the smallest data set, with data collection for only 100 seconds at 20 pA beam current, included only $7.0 \times 10^3$ reconstructed tracks.

Although the beam has finite width, its position is modelled in reconstruction by its central axis, of infinitesimal width. To illustrate, consider the scenario in Figure 8, in which a particle produced near the edge of the beam results in a perfectly-reconstructed track which passes exactly through the original site of reaction. This track will likely pass closer to the central axis at a slightly different depth than the true site of reaction, causing the events in a plane of constant depth, such as the green foil in Figure 8, to be reconstructed in a Gaussian distribution centered about that depth. Modelling a phantom as a series of contiguous foils, each foil at a specific depth, demonstrates how this contribution to distribution shape can obscure the boundaries of various features, including the edges of the phantom itself. Additional contributions to this smearing of interaction vertices may also be due to secondary particles which have scattered in such a way that they appear to have originated from the beam at a different depth (Finck *et al* 2017). While the rate of reconstruction of such false secondary particles is reduced by the fIVI method, these particles and their contributions to the vertex distribution cannot be eliminated entirely.

The field of view of the tracker also influences the shape of the proximal edge of the distribution. For this experiment, the trackers' field of view was centered near the center of the region of interest for range shift testing, with the central axis of the trackers aimed at a BP depth of 30 mm. With the position-sensitive detectors used, detection efficiency is greatest for particles which pass closest to the center of both detectors, and poorer for particles which pass near any edges. Particles produced by events further from the target BP depth will, by necessity, not be able to pass near the center of both detectors, and so suffer from decreased efficiency. This decreased efficiency is visible in the proximal edge of Figure 9, where the number of interaction vertices does not increase immediately upon entering the phantom at a depth of 0 mm, but increases more dramatically at a depth of approximately 10 mm.

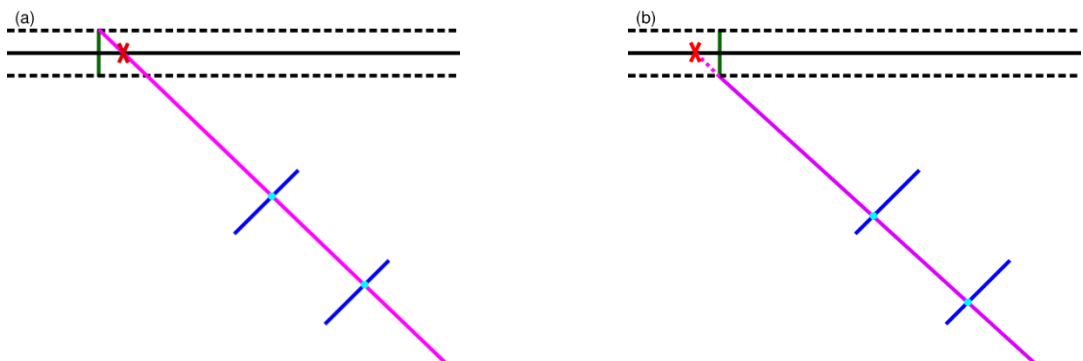

Figure 8: Schematic of the reconstruction process for a single secondary particle. a) A fragmentation reaction occurs at the depth marked with a green vertical line, along the upper edge of the beam (black, dashed), producing a proton (magenta). The proton interacts with two PSDs (blue), and two hits (cyan) are recorded. The resultant proton trajectory follows the same magenta path, but the reconstructed interaction vertex (red cross) appears along the central beam axis (black, solid), producing a longitudinal deviation from the true depth. b) A fragmentation reaction at the same depth, but occurring at the lower edge of the beam. The measured proton trajectory is extended (magenta, dotted) back to the central beam axis, this time producing a longitudinal deviation in the opposite direction to the case in a).

The distal edge of the vertex distribution, also visible in Figure 9, is the feature used for range verification. This edge also has the expected logistic shape, derived from both the beam-width and scattering effects and from the reduced probability of a secondary particle being produced with sufficient energy to reach an external detector as beam energy decreases close to the BP.

### 3.4 RANGE VERIFICATION

The goal of this analysis is to determine the relative shift in BP position for any two measurements, using the characteristic distribution of reconstructed secondary particles emitted in the distal edge region. A representative example comparing two BP depths is shown in Figure 9. The logistic curves are fit to the distal edge regions only (using the definition from section 2.8), although the fit function used in range shift determination is extended to the full plot region. The reconstructed range difference between two depths is determined based on the horizontal translation (in 100 µm increments) required for the best fit of the second fit function to the first, subject to a $\chi^2$ minimization. The proximal (rising) edges of the distributions, once normalized, are very similar in shape. This similarity is expected, as the phantom and detector positions were unchanged between measurements.

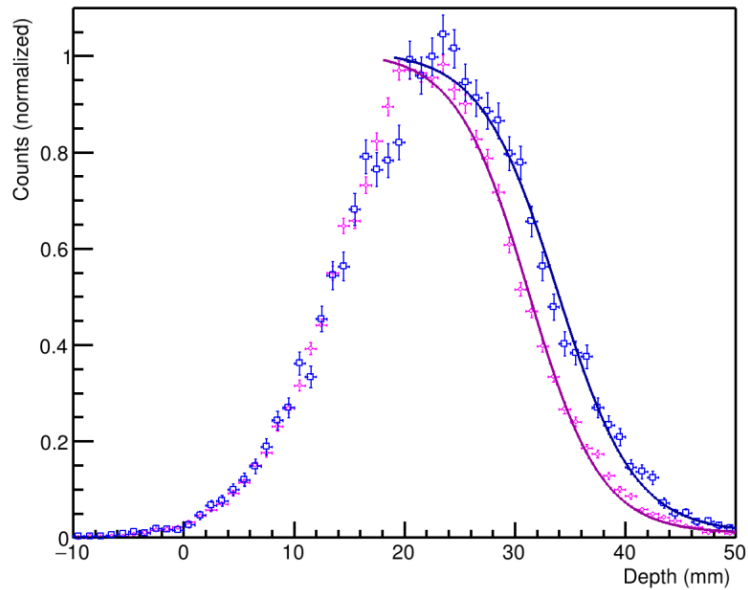

Figure 9: Comparison of experimental vertex distributions at 29.7 mm BP depth (blue squares) and 26.8 mm BP depth (magenta circles). Absolute ranges were calculated by LISE++. Fits to the distal edge region are shown by overlaid curves in the matching colour; both distributions are normalized such that this fit has a height of 1. The $\chi^2$ minimization between the fits is achieved with a 2.8 mm translation of the blue curve, a 100 µm error from the true depth difference of 2.9 mm.

The aggregate results of comparing all measurements are shown in Figure 10. A comparison between two measurements is performed in two ways, with each measurement acting as both the reference data set and the test data set at different times; each of these comparisons is shown as a separate point. Results are similar in magnitude between the two types of comparison, with an average difference of 0.05(9) mm. In this data set, there are three apparent regimes, all with similar slope, but variation in the intercept; the source of these variations is a single offset in the raw data.

Shifts in BP position were accomplished through rotation of a degrader, changing its effective thickness. This effective thickness was controlled remotely by a stepper motor, which was disabled during data collection to reduce electromagnetic noise in the data acquisition system. The single observed offset in Figure 10 is believed to correspond to a skipped rotation, in which the rotation command was sent, but the motor was not enabled, causing an incorrect rotation to be reported. When the skipped position and all subsequent positions are adjusted for the missed 5° rotation, the entire data set is fully correlated. The visibility of these three distinct regimes in the experimental data, allowing identification of a 600 μm offset, serves as an effective demonstration of the precision and robustness of the fIVI algorithm.

The corrected results shown in Figure 11 are well-explained by a linear model, with a $\chi^2$/ndf of 2.11, and a slope approaching 1. Although this linearity with slope 1 is expected from simulation (Hymers and Mücher 2019), this result provides experimental confirmation that fIVI directly measures the true range shift, without requiring any further calibration. Small deviations from unity may derive from the depth-dependence of the exit path length, with lower BP depths requiring a longer exit path to reach the detector, relative to a point with equivalent remaining energy for a deeper Bragg peak. However, the magnitude of this correction is sufficiently small (200 μm over the 6.7 mm region tested) that no patient-specific correction factor would be necessary in converting a reconstructed shift to a true depth shift.

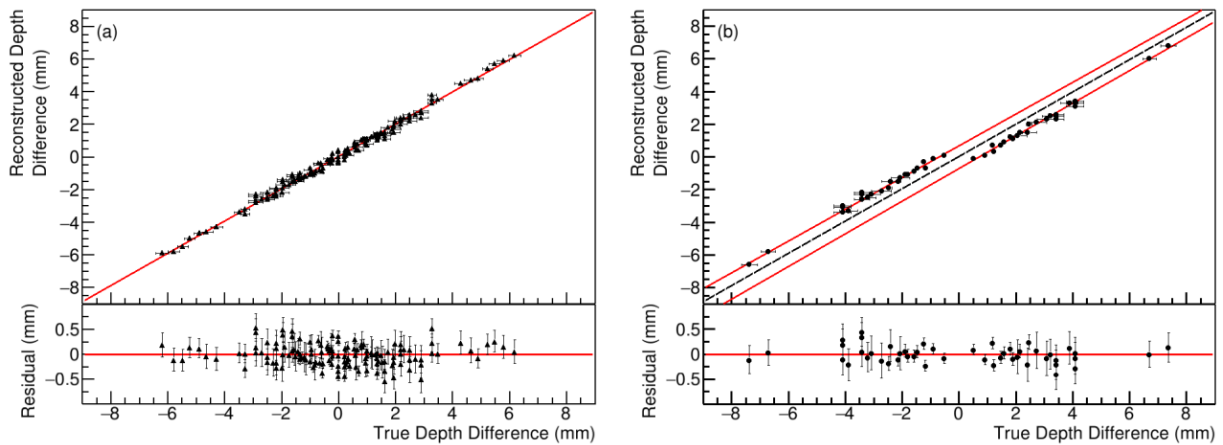

Figure 10: Summary of comparisons as in Figure 9 for all 210 permutations of two Bragg peak depths. a) Comparisons unaffected by the skipped degrader rotation. b) Comparisons affected by the skipped degrader rotation. The central black line repeats the trendline from a), while the two surrounding lines show comparisons offset by approximately 600 μm as a result of this offset. The symmetry of these two lines is derived from whether the affected data point is used as the reference curve or test curve. Residuals are plotted relative to the associated fit function, with a guide line present at a residual of zero.

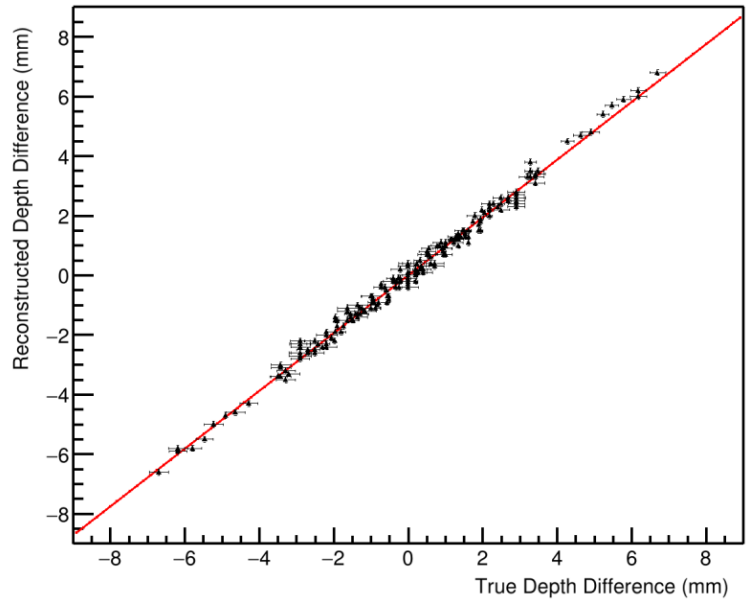

Figure 11: Correction of the two offsets observed in Figure 10. The fit shows a slope of 0.969(5), a nearly one-to-one correspondence between the true depth difference (from LISE++) of two BP positions and the depth difference reconstructed by fIVI.

The residuals in Figure 12 are symmetrically distributed, even when only examining shifts in a single direction from the reference depth. The standard deviation of 220(10) μm indicates that this method achieves sub-millimeter precision, with the greatest deviation between the tested and reconstructed depth differences being less than 700 μm. This precision approaches the physical limit imposed by the beam used for testing, with longitudinal straggling and the effects of a broadened beam imposing uncertainty in BP position on the order of 100 μm; the mean energy and range of the broadened beam incident on the PMMA phantom is used to evaluate the true depth difference. While simulations of monoenergetic beams do produce a steeper distal edge, the steeper edge does not correspond to any improved precision in BP position. This unavoidable uncertainty in BP position, even in a target of known composition, indicates that fIVI measures the range shift from a reference BP depth with precision approaching the physical limit.

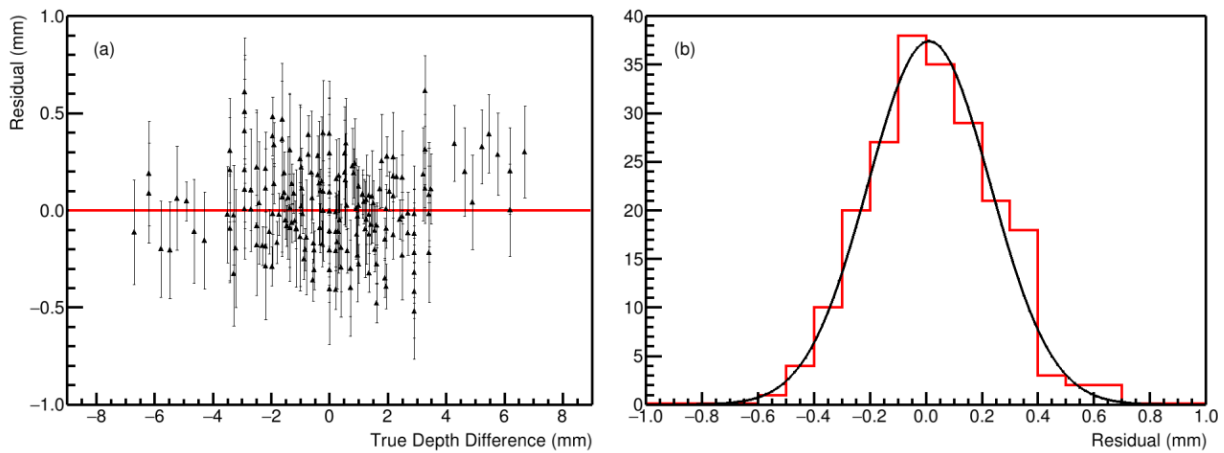

Figure 12: a) Residual plot for the data from Figure 11. In all cases, the error in reconstruction is less than 700 μm. b) Projection of the residual data onto the vertical axis of a), showing a Gaussian distribution with a standard deviation of the mean of 220(10) μm, and equivalent full width at half maximum of 515(25) μm.

# 4 DISCUSSION

## 4.1 RANGE VERIFICATION PERFORMANCE

The filtered interaction vertex imaging method, with its additional filters to reduce the impact of multiple scattering and the use of a full fit and shift method to determine range differences, provides a highly precise method of relative range verification. BP depth differences of less than 1 mm, such as the 600 micron depth difference between BPs at 27.7 mm and 28.3 mm, can be consistently identified and measured in the presented data. This identification is precise, with a standard deviation of the mean of 220(10) µm, performance which has not been achieved in previous studies. This high performance was achieved with a maximum vertex distribution density between 300 vertices mm$^{-1}$ and 900 vertices mm$^{-1}$, which is comparable to previous IVI studies (Henriquet *et al* 2012, Finck *et al* 2017).

The accuracy of the reconstructed depth difference is, as expected, related to the vertex density. Comparisons between two high-density vertex distributions typically match the true depth difference to within 200 µm, while comparisons where one or both vertex distributions are low-density have slightly degraded performance, and typically demonstrate accuracy of 300-400 µm. A comparison between two low-density distributions produces the worst-case result in this data set of a 700 µm inaccuracy in the reconstructed BP depth difference. The vertex distribution density in a clinical setting may depend on patient geometry, such as the exit path length for secondary particles, and individual treatment plans and fractionation schemes, which will affect the number of ions delivered at each depth.

## 4.2 DETECTOR PROPERTIES

One key challenge in IVI is collecting sufficient statistics to produce a robust vertex distribution which can be reliably used for range monitoring. The available ions at each position are defined by the treatment plan and the prescribed dose distribution; any clinically useful RV technique is limited to the number of primary ions delivered at each raster point. Although the results presented here were produced using a larger number of primary ions than would be desirable clinically, they are expected to scale to clinical beams by addressing the efficiency issues of the PSDs and the tracker design, and by increasing the detectors' sensitive area to a similar size as has been investigated clinically (Fischetti *et al* 2020, Toppi *et al* 2021).

From the results presented in section 3.2, it is clear that position-sensitive detectors do not provide the required efficiency at the high count rates experienced in an IVI setup. As the entire sensitive area is involved in every hit, and a relatively long coincidence window is required between hits due to the long charge collection time, even small PSDs such as those used in this study are unlikely to provide the required performance for fIVI. Instead, segmented detectors, which allow faster charge collection as they lack the resistive layer, should be preferred for these applications; the faster charge collection will allow not only higher event rates, but tighter coincidence windows. However, the adequate tracking performance of these PSDs, with a best-case spatial resolution of 200 µm, indicates that the highly-segmented 18.4 to 55 µm square pixels used in previous studies are smaller than required (Gwosch *et al*

2013, Finck *et al* 2017, Reinhart *et al* 2017, Félix-Bautista *et al* 2019). Instead, it should be possible to use a strip-segmented detector of modest pitch, such as 300 µm, which will break the sensitive area of even a large 10 cm × 10 cm detector into segments smaller than the entire 2 cm × 2 cm PSD used in this study. Although the spatial resolution of a 300 µm strip may still be higher than required for maximum reconstruction performance (Fischetti *et al* 2020), maintaining a moderate strip pitch allows reasonable count rates on the order of 10 kHz for each segment, even at high beam intensities of order $10^9$ ion s$^{-1}$.

The presented results also used trackers at a relatively large off-axis angle of 45 degrees, and used the same sensitive area for both the front and rear detector of the tracker. Both of these choices limit the performance of the detection system, as a secondary particle must pass through both detectors to be included in reconstruction, and the reduced flux of secondary particles at a large off-axis angle is compounded by the limits to geometric efficiency of the rear detector. To make best use of both detectors in the tracker, it is desirable to increase the sensitive area of the rear detector, such that both the front and rear detectors provide similar solid angle coverage. For the tested configuration, similar coverage would be achieved by replacing the rear detector in each tracker with a square array of four identical detectors. To compensate for the lower secondary particle flux, the size of all detectors in the tracker could also be increased. Detectors of 10 cm × 10 cm are commonly available, and would provide a significant increase in total detection efficiency while maintaining a manageable size for the overall two-arm design. The total coverage and space requirements for two tracker arms, each with a 20 cm × 20 cm rear sensitive area (as a grid of four 10 cm × 10 cm detectors) is comparable in size to the 30 cm × 30 cm scintillator IVI design successfully integrated into clinical practice by Fischetti *et al* (2020).

Monte Carlo simulations comparing our experimental configuration (with the radial detection efficiency correction suggested by Figure 7) to one which replaces both arms of the tracker with the suggested array of larger segmented detectors increases the number of reconstructed vertices by a factor of 845. This change alone reduces the number of required ions from the $1.29 \times 10^{10}$ delivered by the highest-intensity 50 pA $^{16}O^{8+}$ beam and longest delivery time of 330 s to only $1.52 \times 10^7$. The better-than-expected improvement is attributed to the larger detectors measuring secondary particles exiting the patient at smaller angles relative to the primary beam axis, which are significantly more likely to be produced (Finck *et al* 2017). This simulation does assume full efficiency for the segmented detectors, while an efficiency correction is applied to the PSD output; however, small efficiency differences from simulation may be resolved by making small adjustments in tracker angle to take advantage of the increased secondary particle yield at smaller off-axis angles. As the range shift results presented in Figure 11 include irradiations as short as 100 s and intensities as low as 20 pA, fIVI is expected to remain highly precise, with a worst-case error of less than 700 µm, even for irradiations producing lower numbers of secondary particles. Therefore, a two-arm, 10 × 10 cm$^2$ front and 20 × 20 cm$^2$ rear fIVI setup is expected to identify the range shift between two BP positions with sub-millimeter precision for as few as $5 \times 10^6$ ions delivered to a single depth, validating previous simulations (Hymers and Mücher 2019). This requirement, while insufficient to monitor the position of each raster point in a pencil beam scanned treatment plan, matches the number of ions delivered to each depth in typical treatment plans (Haberer *et al* 1993, Krämer *et al* 2000).

While readouts for strip-segmented detectors are more complex than those for PSDs, significantly fewer channels are required than for pixel-segmented detectors, particularly with the proposed large sensitive areas. The much lower technical requirements for construction of a detector with 300 µm thickness (as in the tested PSDs) and moderate spatial resolution will allow low-cost and highly-effective scaling to the larger sensitive areas required to provide fIVI in clinical settings. Appropriate detectors are already in development in our group, along with accompanying software for rapid readout and online fIVI reconstruction.

## 4.3 CLINICAL APPLICABILITY AND FUTURE WORK

The fIVI method, like all implementations of IVI, is a noninvasive monitoring technique, requiring no implanted markers or injected tracers. Because IVI collects all necessary data during treatment, no additional time is required of the patient for post-treatment verification scans (Henriquet *et al* 2012, Gwosch *et al* 2013, Finck *et al* 2017). These benefits would allow fIVI to be used for online range monitoring of every fraction in a treatment plan.

Although this work has investigated fIVI for use with $^{16}$O beams, the underlying physics of the fragmentation reaction are similar, and previous simulations have demonstrated similar performance with $^{12}$C beams (Hymers and Mücher 2019). While the number of secondary particles reaching detectors may be slightly decreased for the lower-mass, and consequently lower-energy $^{12}$C beams as compared to the studied $^{16}$O, the total number of secondary particles is still sufficient to achieve these results (Finck *et al* 2017), and these relatively small differences in absolute particle number can be compensated with small adjustments in the angle of tracker arms, as discussed in section 4.2. Future investigations with clinical $^{12}$C beams are planned to validate the predicted behaviour.

While the current work has studied offline fIVI reconstruction, there is no technical obstacle to this reconstruction occurring during treatment to provide real-time monitoring. With segmented detectors, the determination of hit positions may be implemented as a lookup table of precomputed values based on detector and strip positions. The reconstruction process is a perfect candidate for parallel processing, and the vectorial nature of the algorithm makes it a strong candidate for GPU acceleration. The determination of range shift can similarly be parallelized by simultaneously computing $\chi^2$ statistics for several shifts of the test distribution, with further speed increases possible by reducing the search area to a width of a few centimeters, centered about the expected range shift. With these speed optimizations, fIVI should be capable of real-time range monitoring, and could even be used as a safety interlock, halting treatment if the beam deviates from the expected or previously-measured depth to an unacceptable degree.

In addition to the intra-fraction monitoring studied in this work, fIVI may also be used for inter-fraction monitoring, to determine if a treatment beam of the same energy is positioned in the same location in the patient as a previous (reference) fraction delivered along the same path. As positions of internal patient structures may change on a much shorter timescale than that of the full treatment plan, the same plan may not always be appropriate for fractions given on different days (Handrack *et al* 2017). Clinical use of fIVI will allow early detection of overall range shifts related to inter-fraction differences with the delivery of only 1-2% of the fraction's total dose, allowing an inappropriately-targeted fraction to be modified or aborted. Furthermore, fIVI provides guidance on the degree to which a BP overranges or underranges a previous fraction, which may be used in adjusting the delivery of the remaining dose in that fraction.

To apply this technique for inter-fraction monitoring, where patient and detector position may vary between fractions, an alignment must be performed to ensure that the comparison of vertex distributions is valid. Any uncertainty in alignment between measurements will directly contribute to uncertainty in the difference in BP depth, so minimizing and measuring this uncertainty will be of great clinical importance. Although the current work has focused on intra-fraction monitoring, where the detectors may be assumed to remain in a fixed position for all depths in the fraction, some of the results are applicable to this inter-fraction alignment as well. For instance, features such as the proximal edge of the vertex distribution can be used for alignment using a similar logistic fit and shift method, to quantify differences in detector positioning relative to the patient. As the proximal edge may be subject to field of view issues for smaller detectors or particularly deep tumours, this same technique might also be applied using inhomogeneities in the patient, if any features significantly affecting the fragmentation reaction cross section lie along the beam path.

This work demonstrates the high accuracy and precision of fIVI for relative RV, comparing two beams delivered at different depths for intra-fraction RV, or the same beam delivered on different days for inter-fraction RV. However, fIVI alone cannot provide absolute RV, which would ensure that a single fraction is delivered to the intended depth. To allow absolute RV, fIVI may be combined with another range verification method, such as a PET scan, to provide a reference measurement at a known BP depth. However, given that these methods also typically rely on Monte Carlo simulations to produce absolute measurements, it should be equally possible to perform these simulations for fIVI directly, with appropriate care given to alignment, as the precision of fIVI for absolute RV will be limited by the precision of this reference measurement. Once a patient-specific reference for fIVI is established, all further fractions may be monitored using fIVI only, providing the benefits of real-time and all-fraction monitoring.

In clinical practice, it is anticipated that the overhead and setup time to align and use fIVI will be minimal. One possible implementation, similar to existing in-beam PET systems, is to integrate the detectors into the treatment gantry, fixing their position relative to the beam axis, and allowing detector positioning to benefit from existing practices for alignment of the beam to the patient. Small deviations in detector alignment between fractions can then be corrected computationally as previously described. If fIVI is used for intra-fraction relative RV only, as modelled in this study, no alignment is necessary, so long as the detectors and patient remain stationary for the duration of treatment. Similarly, the beam axis measurement could be based on the known beam positions prescribed by the treatment plan, and monitored by continual quality assurance of the beamline and steering system.

To pursue clinical applications, further testing under more clinical conditions is required. Now that the fIVI technique has been validated, further validation with clinical beams is necessary, including testing with larger human-scale targets incorporating tissue inhomogeneities, and correspondingly higher beam energies. Using the updated detector system discussed in section 4.2, these tests can be conducted up to clinical beam intensities, and with more stringent controls on the number of ions delivered to each BP position. With these tests, the sensitivity of fIVI will be investigated, to verify that sensitivity and appropriate vertex density remain at lower primary ion counts of $5 \times 10^6$. Testing with inhomogeneous targets will also be required, both to investigate the impacts of these inhomogeneities on the range sensitivity of fIVI, and to validate inhomogeneities as markers to allow alignment corrections for detectors.

# 5 CONCLUSION

The filtered Interaction Vertex Imaging technique is able to determine the difference between two Bragg peak positions from a $^{16}$O beam in a PMMA phantom with sub-millimeter accuracy and precision. This noninvasive technique tracks secondary particles emitted through interactions between a treatment beam and the phantom, with all required data being collected online by thin silicon detectors during treatment. 210 comparisons between Bragg peak positions for a subclinical $^{16}$O beam in the phantom identified the delivered depth difference within 220(10) µm (1σ) of the true value, which has not been achieved by any previous method. As this overall precision is achievable without significant constraints on the reconstruction precision of individual secondary particles, the barrier to clinical implementation of fIVI is lower than other IVI methods which use highly-segmented detectors. However, position-sensitive detectors are not appropriate for this application. Clinical use of fIVI is projected to allow real-time monitoring of relative Bragg peak position for each energy step in a pencil beam scanned treatment plan, as well as inter-fraction monitoring when combined with an offline range verification technique to establish an absolute range reference. Development of larger high-rate segmented detectors is underway, to continue validation of fIVI in more realistic settings, including inhomogeneous human-scale phantoms treated with clinical beam energies and intensities.

# 6 ACKNOWLEDGEMENTS

We acknowledge the support of the CIHR, NSERC, and SSHRC (under Award No. NFRFE-2018-00691), and through the USRA and CGS-M programs. Support was also provided by the Government of Ontario, through an Ontario Graduate Scholarship. This work was made possible by the facilities of the Shared Hierarchical Academic Research Computing Network (SHARCNET: www.sharcnet.ca) and Compute/Calcul Canada. This material is based upon use of NSF's National Superconducting Cyclotron Laboratory which is a major facility fully funded by the National Science Foundation.

# 7 REFERENCES

Agostinelli S, Allison J, Amako K, Apostolakis J, Araujo H, Arce P, Asai M, Axen D, Banerjee S, Barrand G, Behner F, Bellagamba L, Boudreau J, Broglia L, Brunengo A, Burkhardt H, Chauvie S, Chuma J, Chytracek R, Cooperman G, Cosmo G, Degtyarenko P, Dell'Acqua A, Depaola G, Dietrich D, Enami R, Feliciello A, Ferguson C, Fesefeldt H, Folger G, Foppiano F, Forti A, Garelli S, Giani S, Giannitrapani R, Gibin D, Gómez Cadenas J J, González I, Gracia Abril G, Greeniaus G, Greiner W, Grichine V, Grossheim A, Guatelli S, Gumplinger P, Hamatsu R, Hashimoto K, Hasui H, Heikkinen A, Howard A, Ivanchenko V, Johnson A, Jones F W, Kallenbach J, Kanaya N, Kawabata M, Kawabata Y, Kawaguti M, Kelner S, Kent P, Kimura A, Kodama T, Kokoulin R, Kossov M, Kurashige H, Lamanna E, Lampén T, Lara V, Lefebure V, Lei F, Liendl M, Lockman W, Longo F, Magni S, Maire M, Medernach E, Minamimoto K, Mora de Freitas P, Morita Y, Murakami K, Nagamatu M, Nartallo R, Nieminen P, Nishimura T, Ohtsubo K, Okamura M, O'Neale S, Oohata Y, Paech K, Perl J, Pfeiffer A, Pia M G, Ranjard F, Rybin A, Sadilov S, Di Salvo E, Santin G, Sasaki T, et al 2003 Geant4—a simulation toolkit *Nucl. Instrum. Meth. A* **506** 250–303

Amaldi U and Braccini S 2011 Present challenges in hadrontherapy techniques *Eur. Phys. J. Plus* **126** 70

Amaldi U, Hajdas W, Iliescu S, Malakhov N, Samarati J, Sauli F and Watts D 2010 Advanced Quality Assurance for CNAO *Nucl. Instrum. Meth. A* **617** 248–9

Amaldi U and Kraft G 2005 Radiotherapy with beams of carbon ions *Rep. Prog. Phys.* **68** 1861–82

Ammazzalorso F, Graef S, Weber U, Wittig A, Engenhart-Cabillic R and Jelen U 2014 Dosimetric consequences of intrafraction prostate motion in scanned ion beam radiotherapy *Radiother. Oncol.* **112** 100–5

Barton M B 2014 Estimating the demand for radiotherapy from the evidence: A review of changes from 2003 to 2012 *Radiother. Oncol.* **112** 140–4

Bray F, Ferlay J, Soerjomataram I, Siegel R L, Torre L A and Jemal A 2018 Global cancer statistics 2018: GLOBOCAN estimates of incidence and mortality worldwide for 36 cancers in 185 countries *CA Cancer J. Clin.* **68** 394–424

Brenner D R, Weir H K, Demers A A, Ellison L F, Louzado C, Shaw A, Turner D, Woods R R and Smith L M 2020 Projected estimates of cancer in Canada in 2020 *Can. Med. Assoc. J.* **192** E199–205

Brun R and Rademakers F 1997 ROOT — An object oriented data analysis framework *Nucl. Instrum. Meth. A* **389** 81–6

Enghardt W, Parodi K, Crespo P, Fiedler F, Pawelke J and Pönisch F 2004 Dose Quantification from In-Beam Positron Emission Tomography *Radiother. Oncol.* **73** S96–8

Félix-Bautista R, Gehrke T, Ghesquière-Diérickx L, Reimold M, Amato C, Turecek D, Jakubek J, Ellerbrock M and Martišíková M 2019 Experimental verification of a non-invasive method to monitor the lateral pencil beam position in an anthropomorphic phantom for carbon-ion radiotherapy *Phys. Med. Biol.* **64** 175019

Finck C, Karakaya Y, Reithinger V, Rescigno R, Baudot J, Constanzo J, Juliani D, Krimmer J, Rinaldi I, Rousseau M, Testa E, Vanstalle M and Ray C 2017 Study for online range monitoring with the interaction vertex imaging method *Phys. Med. Biol.* **62** 9220–39

Fischetti M, Baroni G, Battistoni G, Bisogni G, Cerello P, Ciocca M, De Maria P, De Simoni M, Di Lullo B, Donetti M, Dong Y, Embriaco A, Ferrero V, Fiorina E, Franciosini G, Galante F, Kraan A, Luongo C, Magi M, Mancini-Terracciano C, Marafini M, Malekzadeh E, Mattei I, Mazzoni E, Mirabelli R, Mirandola A, Morrocchi M, Muraro S, Patera V, Pennazio F, Schiavi A, Sciubba A, Solfaroli Camillocci E, Sportelli G, Tampellini S, Toppi M, Traini G, Valle S M, Vischioni B, Vitolo V and Sarti A 2020 Inter-fractional monitoring of $^{12}$C ions treatments: results from a clinical trial at the CNAO facility *Sci Rep* **10** 20735

Gade A and Sherrill B M 2016 NSCL and FRIB at Michigan State University: Nuclear science at the limits of stability *Phys. Scr.* **91** 053003

Gwosch K, Hartmann B, Jakubek J, Granja C, Soukup P, Jäkel O and Martišíková M 2013 Non-invasive monitoring of therapeutic carbon ion beams in a homogeneous phantom by tracking of secondary ions *Phys. Med. Biol.* **58** 3755–73

Haberer Th, Becher W, Schardt D and Kraft G 1993 Magnetic scanning system for heavy ion therapy *Nucl. Instrum. Meth. A* **330** 296–305

Handrack J, Tessonnier T, Chen W, Liebl J, Debus J, Bauer J and Parodi K 2017 Sensitivity of post treatment positron emission tomography/computed tomography to detect inter-fractional range variations in scanned ion beam therapy *Acta Oncol.* **56** 1451–8

Henriquet P, Testa E, Chevallier M, Dauvergne D, Dedes G, Freud N, Krimmer J, Létang J M, Ray C, Richard M-H and Sauli F 2012 Interaction vertex imaging (IVI) for carbon ion therapy monitoring: a feasibility study *Phys. Med. Biol.* **57** 4655–69

Hymers D and Mücher D 2019 Monte Carlo investigation of sub-millimeter range verification in carbon ion radiation therapy using interaction vertex imaging *Biomed. Phys. Eng. Express* **5** 045025

Jäkel O, Krämer M, Karger C P and Debus J 2001 Treatment planning for heavy ion radiotherapy: clinical implementation and application *Phys. Med. Biol.* **46** 1101–16

Krämer M, Jäkel O, Haberer T, Kraft G, Schardt D and Weber U 2000 Treatment planning for heavy-ion radiotherapy: physical beam model and dose optimization *Phys. Med. Biol.* **45** 3299

Krimmer J, Dauvergne D, Létang J M and Testa É 2018 Prompt-gamma monitoring in hadrontherapy: A review *Nucl. Instrum. Meth. A* **878** 58–73

Ladbury R, Reed R A, Marshall P, LaBel K A, Anantaraman R, Fox R, Sanderson D P, Stolz A, Yurkon J, Zeller A F and Stetson J W 2004 Performance of the high-energy single-event effects test Facility (SEETF) at Michigan State university's national Superconducting Cyclotron laboratory (NSCL) *IEEE Trans. Nucl. Sci.* **51** 3664–8

Laprie A, Hu Y, Alapetite C, Carrie C, Habrand J-L, Bolle S, Bondiau P-Y, Ducassou A, Huchet A, Bertozzi A-I, Perel Y, Moyal É and Balosso J 2015 Paediatric brain tumours: A review of radiotherapy, state of the art and challenges for the future regarding protontherapy and carbontherapy *Cancer Radiother.* **19** 775–89

Magalhaes Martins P, Dal Bello R, Ackermann B, Brons S, Hermann G, Kihm T and Seco J 2020 PIBS: Proton and ion beam spectroscopy for in vivo measurements of oxygen, carbon, and calcium concentrations in the human body *Sci. Rep.* **10** 7007

Parodi K and Polf J C 2018 *In vivo* range verification in particle therapy *Med. Phys.* **45** e1036–50

Pennazio F, Battistoni G, Bisogni M G, Camarlinghi N, Ferrari A, Ferrero V, Fiorina E, Morrocchi M, Sala P, Sportelli G, Wheadon R and Cerello P 2018 Carbon ions beam therapy monitoring with the INSIDE in-beam PET *Phys. Med. Biol.* **63** 145018

Reinhart A M, Spindeldreier C K, Jakubek J and Martišíková M 2017 Three dimensional reconstruction of therapeutic carbon ion beams in phantoms using single secondary ion tracks *Phys. Med. Biol.* **62** 4884–96

Roth G A, Abate D, Abate K H, Abay S M, Abbafati C, Abbasi N, Abbastabar H, Abd-Allah F, Abdela J, Abdelalim A, Abdollahpour I, Abdulkader R S, Abebe H T, Abebe M, Abebe Z, Abejie A N, Abera S

F, Abil O Z, Abraha H N, Abrham A R, Abu-Raddad L J, Accrombessi M M K, Acharya D, Adamu A A, Adebayo O M, Adedoyin R A, Adekanmbi V, Adetokunboh O O, Adhena B M, Adib M G, Admasie A, Afshin A, Agarwal G, Agesa K M, Agrawal A, Agrawal S, Ahmadi A, Ahmadi M, Ahmed M B, Ahmed S, Aichour A N, Aichour I, Aichour M T E, Akbari M E, Akinyemi R O, Akseer N, Al-Aly Z, Al-Eyadhy A, Al-Raddadi R M, Alahdab F, Alam K, Alam T, Alebel A, Alene K A, Alijanzadeh M, Alizadeh-Navaei R, Aljunid S M, Alkerwi A, Alla F, Allebeck P, Alonso J, Altirkawi K, Alvis-Guzman N, Amare A T, Aminde L N, Amini E, Ammar W, Amoako Y A, Anber N H, Andrei C L, Androudi S, Animut M D, Anjomshoa M, Ansari H, Ansha M G, Antonio C A T, Anwari P, Aremu O, Ärnlöv J, Arora A, Arora M, Artaman A, Aryal K K, Asayesh H, Asfaw E T, Ataro Z, Atique S, Atre S R, Ausloos M, Avokpaho E F G A, Awasthi A, Quintanilla B P A, Ayele Y, Ayer R, Azzopardi P S, Babazadeh A, Bacha U, Badali H, et al 2018 Global, regional, and national age-sex-specific mortality for 282 causes of death in 195 countries and territories, 1980–2017: a systematic analysis for the Global Burden of Disease Study 2017 *Lancet* **392** 1736–88

Scifoni E, Tinganelli W, Weyrather W K, Durante M, Maier A and Krämer M 2013 Including oxygen enhancement ratio in ion beam treatment planning: model implementation and experimental verification *Phys. Med. Biol.* **58** 3871–95

Soisson S N, Stein B C, May L W, Dienhoffer R Q, Jandel M, Souliotis G A, Shetty D V, Galanopoulos S, Keksis A L, Wuenschel S, Kohley Z, Yennello S J, Bullough M A, Greenwood N M, Walsh S M and Wilburn C D 2010 A dual-axis dual-lateral position-sensitive detector for charged particle detection *Nuclear Instruments and Methods in Physics Research Section A: Accelerators, Spectrometers, Detectors and Associated Equipment* **613** 240–4

Sokol O, Scifoni E, Tinganelli W, Kraft-Weyrather W, Wiedemann J, Maier A, Boscolo D, Friedrich T, Brons S, Durante M and Krämer M 2017 Oxygen beams for therapy: advanced biological treatment planning and experimental verification *Phys. Med. Biol.* **62** 7798–813

Tarasov O B and Bazin D 2016 LISE++: Exotic beam production with fragment separators and their design *Nucl. Instrum. Meth. B* **376** 185–7

Tessonnier T, Mairani A, Brons S, Haberer T, Debus J and Parodi K 2017 Experimental dosimetric comparison of $^1$H, $^4$He, $^{12}$C and $^{16}$O scanned ion beams *Phys. Med. Biol.* **62** 3958–82

Toppi M, Baroni G, Battistoni G, Bisogni M G, Cerello P, Ciocca M, De Maria P, De Simoni M, Donetti M, Dong Y, Embriaco A, Ferrero V, Fiorina E, Fischetti M, Franciosini G, Kraan A C, Luongo C, Malekzadeh E, Magi M, Mancini-Terracciano C, Marafini M, Mattei I, Mazzoni E, Mirabelli R, Mirandola A, Morrocchi M, Muraro S, Patera V, Pennazio F, Schiavi A, Sciubba A, Solfaroli-Camillocci E, Sportelli G, Tampellini S, Traini G, Valle S M, Vischioni B, Vitolo V and Sarti A 2021 Monitoring Carbon Ion Beams Transverse Position Detecting Charged Secondary Fragments: Results From Patient Treatment Performed at CNAO *Front. Oncol.* **11** 601784

Tsujii H, Mizoe J, Kamada T, Baba M, Kato S, Kato H, Tsuji H, Yamada S, Yasuda S, Ohno T, Yanagi T, Hasegawa A, Sugawara T, Ezawa H, Kandatsu S, Yoshikawa K, Kishimoto R and Miyamoto T 2004 Overview of clinical experiences on carbon ion radiotherapy at NIRS *Radiother. Oncol.* **73** S41–9

Tyldesley S, Delaney G, Foroudi F, Barbera L, Kerba M and Mackillop W 2011 Estimating the Need for Radiotherapy for Patients With Prostate, Breast, and Lung Cancers: Verification of Model

Estimates of Need With Radiotherapy Utilization Data From British Columbia *Int. J. Radiat. Oncol. Biol. Phys.* **79** 1507–15

Ziegler J F 2004 SRIM-2003 *Nucl. Instrum. Meth. B* **219–220** 1027–36